\documentclass[final,5p,times,twocolumn]{elsarticle}
\biboptions{sort&compress}

\usepackage{amssymb}
\usepackage{lipsum}  
\usepackage{placeins}
\usepackage{subcaption}
\usepackage{overpic}
\usepackage{url}
\usepackage[pdftex, pdftitle={Article}, pdfauthor={Author},colorlinks = true,allcolors = blue]{hyperref}
\usepackage{soul}
\usepackage{calrsfs}
\DeclareMathAlphabet{\pazocal}{OMS}{zplm}{m}{n}

\newcommand{\Lb}{\pazocal{L}}
\usepackage{lineno}
\journal{NIMA}

\begin{document}

\affiliation[UCR]{Department of Physics and Astronomy, University of California, Riverside, CA 92521, USA}
\affiliation[LLNLNACS]{Nuclear and Chemical Science Division, Lawrence Livermore National Laboratory, Livermore, CA 94550}
\affiliation[LLNLCED]{Computational Engineering Division, Lawrence Livermore National Laboratory, Livermore CA 94550}
\affiliation[LBNL]{Physics Division, Lawrence Berkeley National Laboratory, Berkeley, CA 94720, USA}
\affiliation[BIDS]{Berkeley Institute for Data Science, University of California, Berkeley, CA 94720, USA}

\author[UCR]{Ryan Milton}
\author[UCR]{Sebouh J. Paul}
\author[UCR]{Barak Schmookler}
\author[UCR]{Miguel Arratia}
\author[LLNLCED]{Piyush Karande}
\author[LLNLNACS]{Aaron Angerami}
\author[LBNL]{Fernando Torales Acosta}
\author[LBNL,BIDS]{Benjamin Nachman}

\title{Design and Simulation of a SiPM-on-Tile ZDC for the future EIC,\\
and its Performance with Graph Neural Networks}

\begin{abstract}
We present a design for a high-granularity zero-degree calorimeter (ZDC) for the upcoming Electron-Ion Collider (EIC). The design uses SiPM-on-tile technology and features a novel staggered-layer arrangement that improves spatial resolution. To fully leverage the design's high granularity and non-trivial geometry, we employ graph neural networks (GNNs) for energy and angle regression as well as signal classification. The GNN-boosted performance metrics meet, and in some cases, significantly surpass the requirements set in the report on  science requirements and detector requirements for the EIC (Yellow Report), laying the groundwork for enhanced measurements that will facilitate a wide physics program. Our studies show that GNNs can significantly enhance the performance of high-granularity CALICE-style calorimeters by automating and optimizing the software compensation algorithms required for these systems. This improvement holds true even in the case of complicated geometries that pose challenges for image-based AI/ML methods.
\end{abstract}

\maketitle
\section{Introduction}
\label{sec:outline}
The Zero-Degree Calorimeter (ZDC) at the future Electron-Ion Collider (EIC)~\cite{Accardi:2012qut} will measure high-energy neutral hadrons and photons, and will be used to reconstruct neutral pion decays to support a comprehensive physics program with electron-proton ($ep$) and electron-nucleus ($eA$) collisions~\cite{EICYR}. For example, the ZDC will be used to measure meson structure with deep exclusive meson production, \textit{i.e.}, $ep\to e\pi^{+}n$ and $ep\to eK^{+}\Lambda (n\pi^{0})$, which generate a neutron or lambda near the beam direction. In these reactions, the ZDC's position resolution plays a crucial role in reconstructing the momentum-transfer variable $t$~\cite{EICYR, Bylinkin:2022rxd}. In another application, the ZDC will measure spectator nucleons in scattering off light ions~\cite{Friscic:2021oti, Jentsch:2021qdp}, or off heavy ions~\cite{Zheng:2014cha, Chang:2022hkt}, where energy linearity plays an important role. Furthermore, the ZDC granularity must possess the capability to distinguish between photons and neutral pions produced in $u$-channel backward reactions, such as deeply virtual $\pi^{0}$ production ($ep\to e\pi^{0}p$) and deeply virtual Compton scattering ($ep\to e\gamma p$)~\cite{Sweger:2023bmx}.

The ZDC performance requirements stated in section 14.5.2 of the report on the science requirements and detector concepts for the EIC (Yellow report, YR)~\cite{EICYR} include a single-hadron energy resolution better than 50\%$/\sqrt{E}$ and an angular resolution better than 3 mrad$/\sqrt{E}$. The required energy range extends up to the beam energy, which will be a maximum of 275 GeV for protons. In $eA$ collisions at maximum energy, the average energy per nucleon for lead-208 nuclei is approximately 110 GeV.

As per the ECCE baseline design for the EIC project detector~\cite{ECCE,ECCE_CALO}, the ZDC incorporates two sampling calorimeter sections: a 12-layer lead/silicon-pad section (21 cm) and a 30-layer lead/scintillator section~\cite{ECCE_CALO}, resulting in a depth of 7$\lambda$. The ZDC's location is at about $z=35$ m and  nominally covers $\theta<5.5$ mrad ($\eta>6$). This design uses lead absorbers with an approximately 4 to 1 lead-to-scintillator thickness ratio to achieve a compensated response (\textit{i.e.}, a similar response for electromagnetic and hadronic showers). This approach illustrates ``hardware compensation'', which improves energy resolution by minimizing the impact of shower-to-shower fluctuations on the electromagnetic fraction of hadronic showers.

An alternative iron-scintillator design offers the advantage of a self-supporting iron structure, easing assembly and maintenance in the ZDC's limited space. Additionally, employing an iron absorber could reduce neutron production by a factor of about 4 compared to a lead absorber~\cite{Wigmans}, thereby mitigating radiation damage in the SiPMs~\footnote{The neutron fluence expected in the ZDC region is projected to be less than $10^{12}$ 1-MeV neutron equivalent per cm$^2$ per year at peak luminosity ($10^{34}$ cm$^{-2}$s$^{-1}$)~\cite{Doses}. This may necessitate mitigation measures in the design, such as employing annealing of SiPMs in between runs~\cite{Garutti:2018hfu}. 
}. A limitation, however, is that this design cannot be compensated at the hardware level with any practical absorber-to-scintillator thickness ratio~\cite{Wigmans:217893}.

The non-compensated nature of iron-scintillator calorimeters can be corrected with ``software compensation'' algorithms, which exploit the shower-cell topology to re-weight electromagnetic and hadronic sub-showers, \textit{e.g.}, Refs.~\cite{ParticleFlowReview, ATLAS:2016krp,CALICE:2012eac}. 
The CALICE collaboration has tested these algorithms with their iron-scintillator SiPM-on-tile prototypes (\textit{e.g.}, Ref.~\cite{CALICE:2022uwn}).

The advent of  modern machine-learning techniques, such as deep learning, offers the prospect of fully exploiting the power of high-granularity calorimetry and software-compensation techniques in an automated, efficient, and optimal way. Several works have shown these, mostly from studies at the LHC~\cite{deOliveira:2018hva,CERN-LHCC-2017-023,ATL-PHYS-PUB-2020-018,Neubuser:2021uui,Akchurin:2021afn,Akchurin:2021ahx,ATL-PHYS-PUB-2022-040,Qasim:2022rww,Kieseler:2021jxc}. In particular, graph neural networks (GNNs), see \textit{e.g.}, Ref.~\cite{Shlomi:2020gdn}, offer a promising approach to handle energy regression for calorimeter showers, even with complex geometries, providing improvements over traditional (non-AI/ML) software compensation methods.

Examples of GNN work for software compensation include work by the ATLAS collaboration~\cite{ATL-PHYS-PUB-2022-040}, by us in an EIC context in Ref.~\cite{codesign}, and more recently by the CALICE collaboration~\cite{CALICE:2024imr}. Concurrently, the GNNs are capable of classifying events by particle species, which is needed for several EIC applications such as to distinguish single photons from $\pi^{0}$ decays or neutrons.

In this paper, we present the design of a high-granularity ZDC for the EIC based on SiPM-on-tile technology and the iron-scintillator design. We highlight this design for its high energy and angular resolution, leveraging its potential through GNNs. Section~\ref{sec:design} showcases the detector design. The simulations and regression approach are described in Section~\ref{sec:reconstruction}, while the performance is outlined in Section~\ref{sec:performance}. Finally, the summary and conclusions are presented in Section~\ref{sec:summary}.

\section{Design}
\label{sec:design}
\begin{figure}[h!]
    \centering
    \includegraphics[width=0.495\textwidth, trim={5cm 0 0 0}, clip]{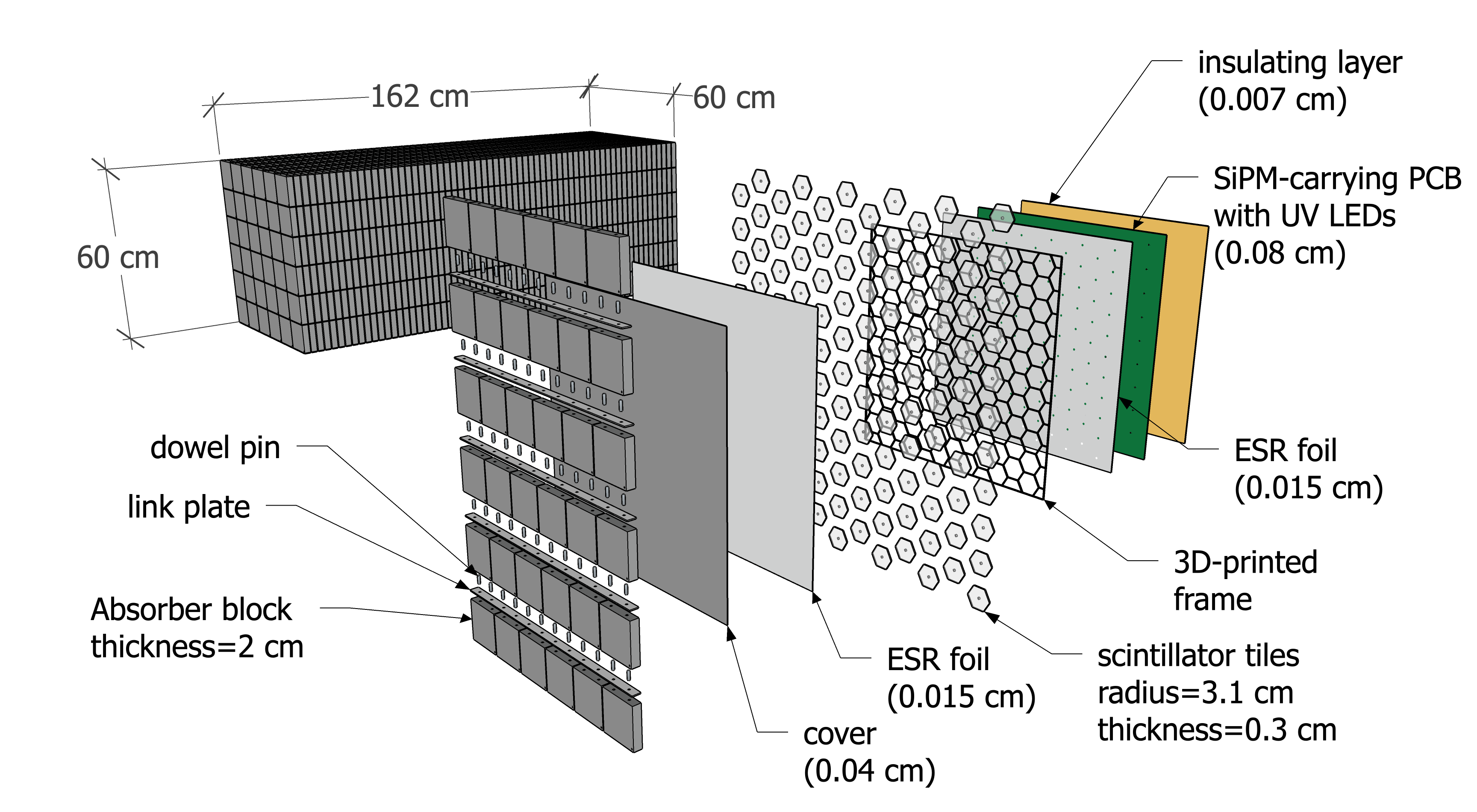}
    \caption{Foreground: exploded view of a ZDC layer.  Background:  the entire ZDC with dimensions.}
    \label{fig:explode_view}
\end{figure}

Figure~\ref{fig:explode_view} displays our design for the ZDC, which is based on the SiPM-on-tile approach~\cite{CALICE:2022uwn}, following a similar approach to that of Ref.~\cite{Insert}. 
The sampling structure consists of iron absorber layers and scintillator layers, which are read out with SiPMs.
\FloatBarrier

The overall dimensions are $60\times60\times162$ cm$^{3}$, driven by limits imposed by the EIC beamlines. The self-supporting iron absorber structure is based on steel blocks with dimensions of $96\mathrm{mm}\times98\mathrm{mm}\times20\mathrm{mm}$, which are identical to the ones currently used in the STAR hadronic calorimeter~\cite{Tsai:2015bna,Aschenauer:2016our}, and that will be accessible for reuse in EIC experiments. 

The scintillator cells will be 25 cm$^{2}$ hexagonal tiles with thickness of 3 mm, featuring a dimple at the center for coupling to a SiPM (Hamamatsu S14160-3015PS). The cells will be made from either machined EJ-200 scintillator or by an injection-molding technique. The scintillator cells could be positioned within 3D-printed plastic frames~\cite{InsertJINST}, which will then be interleaved between reflective foils (ESR by 3M), or by wrapping tiles in ESR like the CALICE design~\cite{CALICE:2022uwn}. 

\begin{figure}[h!]
    \centering
    \includegraphics[width=0.35\textwidth]{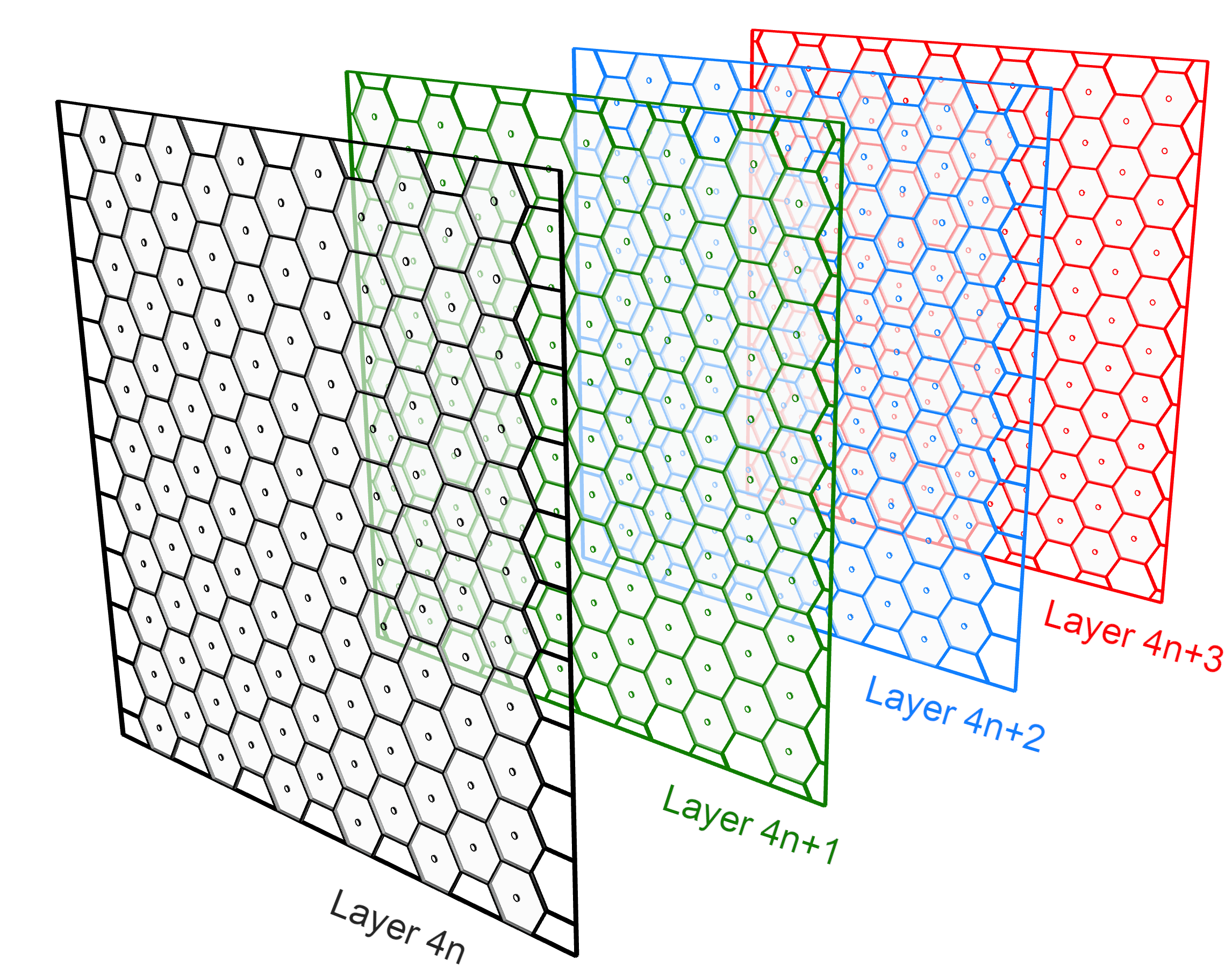}
    \includegraphics[width=0.3\textwidth]{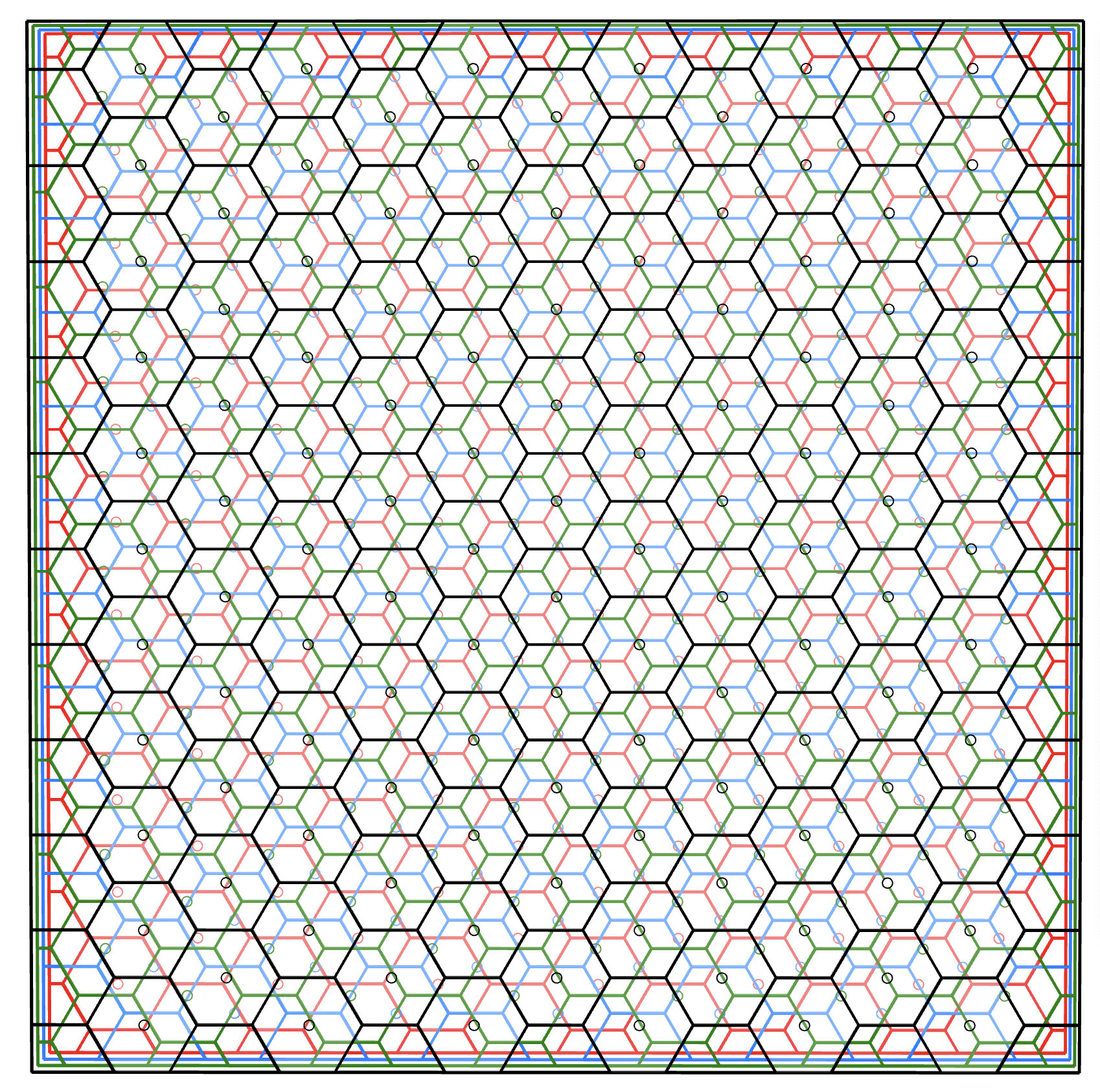}
    \caption{Top: Four scintillator layers with different offsets for the tile positions, with an exaggerated spacing between them.  Bottom: 3D rendering of the same four layers, as viewed head on. Each colored layer shows a different cell layout.}
    \label{fig:overlap}
\end{figure}

Following Ref.~\cite{hexplit}, the layers will cycle through four different scintillator layouts in order to stagger them, as shown in Fig.~\ref{fig:overlap}. This staggering approach enhances the position resolution of the ZDC.  In this four-layer cyclical layout, referred to as ``H4'' in Ref.~\cite{hexplit}, the overlap between the cells in four consecutive layers defines rhombus-shaped subcells that have area equal to 1/12 of that of the hexagonal cells.  

\section{Simulation}
\label{sec:reconstruction}
The \textsc{Geant4}~\cite{GEANT4:2002zbu} (v11.0.p2) simulations of the ZDC geometry were implemented in the \textsc{DD4HEP} framework~\cite{Frank:2014zya} using the FTFP\_BERT physics list. No noise was included in the simulation. Auxiliary simulations of muons and electrons were also generated to define the hit-energy MIP unit and electromagnetic scale of the calorimeter, respectively. We performed studies both with the staggered layout and an unstaggered layout for reference.    

Hits were reconstructed assuming a 15-bit ADC\footnote{The ASIC for the EIC calorimeters will be based on the HGROC3 ASIC developed for the CMS HGAL, which provides an effective dynamic range of 16 bits~\cite{Dulucq:2797658}.} with no noise and a dynamic range of 800 MeV. Only reconstructed hit energies above $E > 0.5$ MIP, with 1 MIP $=$0.5 MeV estimated from muon simulations, and a hit time of $t < 275$ ns were considered for further analysis.  

We defined the reconstructed shower energy at the electromagnetic scale, which was determined using electron simulations. These simulations were used to estimate a sampling fraction of 2.1\%.

A standard reference is defined with the simplest energy regression algorithm, referred to as the ``strawman'', which defines the reconstructed energy as the sum of the energies in cells that pass both the time and energy cut, divided by the sampling fraction obtained with electron simulations. For hadronic showers, this algorithm produces an energy scale that was not unity and exhibited deviations from linearity of 1-2\% when fit with a linear function. These behaviors, as will be shown in Fig.~\ref{fig:energy_resolution}, are expected due to the non-compensated nature of the calorimeter.

As a reference for the position (angle) reconstruction, we used two algorithms: first, a log-weighted center-of-gravity reconstruction (``baseline''), and second, a modified version of this algorithm known as ``HEXPLIT''~\cite{hexplit} which takes advantage of the overlapping cells.  In HEXPLIT, subcells are defined by the overlap of cells with those in neighboring (and next-to-neighboring) layers, and the relative energy contributions in each subcell are estimated using the energy in the overlapping cells in these neighboring and next-to-neighboring layers.  The log-weighted center-of-gravity reconstruction of the shower position is then performed on the subcells instead of on hits in individual cells.  This algorithm is described in further detail in Ref.~\cite{hexplit}.

\subsection{ML-based reconstruction}
In addition to the strawman reconstruction algorithm, an ML-based reconstruction using GNNs~\cite{codesign, ATL-PHYS-PUB-2022-040} was also used. This was done using the Graph Nets library~\cite{GraphNets} in TensorFlow~\cite{tensorflow2015-whitepaper}. For training and test data, we used single-particle simulations of neutrons, $\pi^0$s, and photons with the ``H4" staggering approach for the ZDC's readout. We also used multiple-particle simulations of neutrons with the same configuration.

The calorimeter data for each event was represented as a graph, with each node containing the energies and positions of cells that pass the energy and timing cuts. The nodes were interconnected using a set of edges, which represent the set of the ten nearest neighbors of a cell. A global node containing the summed node energies divided by the sampling fraction was also included in the graph. All energies in the graph were normalized via a z-score normalization using the mean and standard deviation from a subset of events. 

A dense neural network composed of four dense layers with 64 nodes each was trained to predict the generated energy and polar angle, $E_{\rm truth}$ and $\theta_{\rm truth}$, respectively. Each dense layer used the Rectified Linear Unit (ReLU) activation function~\cite{relu} and He-normal initialization \cite{HeUniform}. The model was trained with a batch size of 256 calorimeter showers for 70 epochs, using the Adam optimizer~\cite{adam}. The learning rate was initialized to $1e^{-3}$ and was halved every 5 epochs to a minimum of $1e^{-6}$. 
\subsubsection{Neutrons}
To quantify the single-neutron performance of the ZDC, we generated single-neutron events with energies in the range of 10 GeV--300 GeV, polar angles in the range $0 < \theta < 5$ mrad, and azimuthal angles in the range $0 < \phi < 360^{\circ}$. We trained a model using 750k events as training data and 250k for validation data, with each event energy sampled from a continuous log-uniform distribution. Mean absolute error loss was used, with equal weights given to energy and $\theta$. For testing, we used a separate set of simulated events at discrete energies between 10 GeV and 300 GeV, with a total of 500k events. While the model was trained with data having up to $\theta = 5$ mrad, the test data was limited to  $0 < \theta < 4$ mrad, which is the fiducial acceptance of ZDC.

We also explored multiple-neutron events, which are expected in $eA$ collisions, to test the energy linearity of the ZDC. A model was trained to predict the energy of events with varying numbers of neutrons, $N_{n}$. We trained the model on a mixture of the two types of events the ZDC will measure. 

The first data set contained 1 million events with a random number of neutrons between 1 and 10, each with an energy 10 GeV--200 GeV, sampled from a continuous log-uniform distribution. The second set had 0.5 million events with a random number of neutrons between 1 and 10, each with an energy sampled from a Gaussian distribution of $\mu = 100$ GeV and $\sigma = 5$ GeV. The value of 5 GeV represents the expected smearing caused by nuclear effects in $eA$ collisions, as estimated using the \textsc{BeAgle} event generator~\cite{Chang:2022hkt}. Each neutron in these data sets had random $\theta$ and $\phi$, sampled uniformly in the ranges $0 < \theta < 5$ mrad and $0 < \phi < 360^{\circ}$, respectively. 

The loss function was again the mean absolute error, with $E_{\rm truth}$ being the sum of the generated neutron energies. We trained the model for 100 epochs for the multiple-neutron events, but the loss did not decrease after 70 epochs. For the test data, we used 0.5 million events with 1--10 neutrons, each with an energy of exactly 100 GeV, $0 < \theta < 4$ mrad, and $0 < \phi < 360^{\circ}$. 

The approximate maximum number of neutrons that will hit the ZDC in the ``most-central'' collisions is about 50~\cite{Chang:2022hkt}. However, for this work, we focused on the range 1--10, which captures the bulk of the total $eA$ cross-section and is most relevant for studies such as Ref.~\cite{ALICE:2022iqi}.

\subsubsection{$\pi^{0}/\gamma$}
To investigate $\pi^0/\gamma$ separation, the model was extended to predict $E_{\rm truth}$, $\theta_{\rm truth}$, and the particle type. The modified loss function follows from Ref.~\cite{ATL-PHYS-PUB-2022-040}:
\begin{equation}
    \Lb= (1-\alpha)\Lb_{\rm classification} + \alpha \Lb_{\rm regression},
    \label{eqn:loss}
\end{equation}
where $\Lb_{\rm classification}$ is the binary cross-entropy loss, $\Lb_{\rm regression}$ is the mean absolute error loss, and $\alpha$ is a hyperparameter specifying the importance of classification versus regression. $\alpha$ was set to 0.75 for these studies. Energy and $\theta$ were given equal weights in $\Lb_{\rm regression}$.

The model's output for particle type classification was converted to the probability of an event's incident particle being a $\pi^0$ using the sigmoid function. Since the sigmoid function returns a continuous value between 0 and 1, we invoked a classification cut of $0.3$ -- below which, we called the model's classification a $\gamma$ and above which, the output was a $\pi^0$. 

We used 880k single-$\pi^0$ and 880k single-$\gamma$ events for training the model and 300k for each particle type during validation. The model was trained with a random order of $\pi^{0}$ and $\gamma$ events. The data were again generated with a log-uniform distribution of energies and a polar angle range of $0 < \theta < 4$ mrad. The model was tested using a separate set of simulated events at discrete energies, with 500k events for both $\pi^0$ and $\gamma$ events.

\section{Performance}
\label{sec:performance}
\subsection{Single-neutron energy resolution}
\begin{figure}[h!]
    \centering
    \includegraphics[width=\columnwidth]{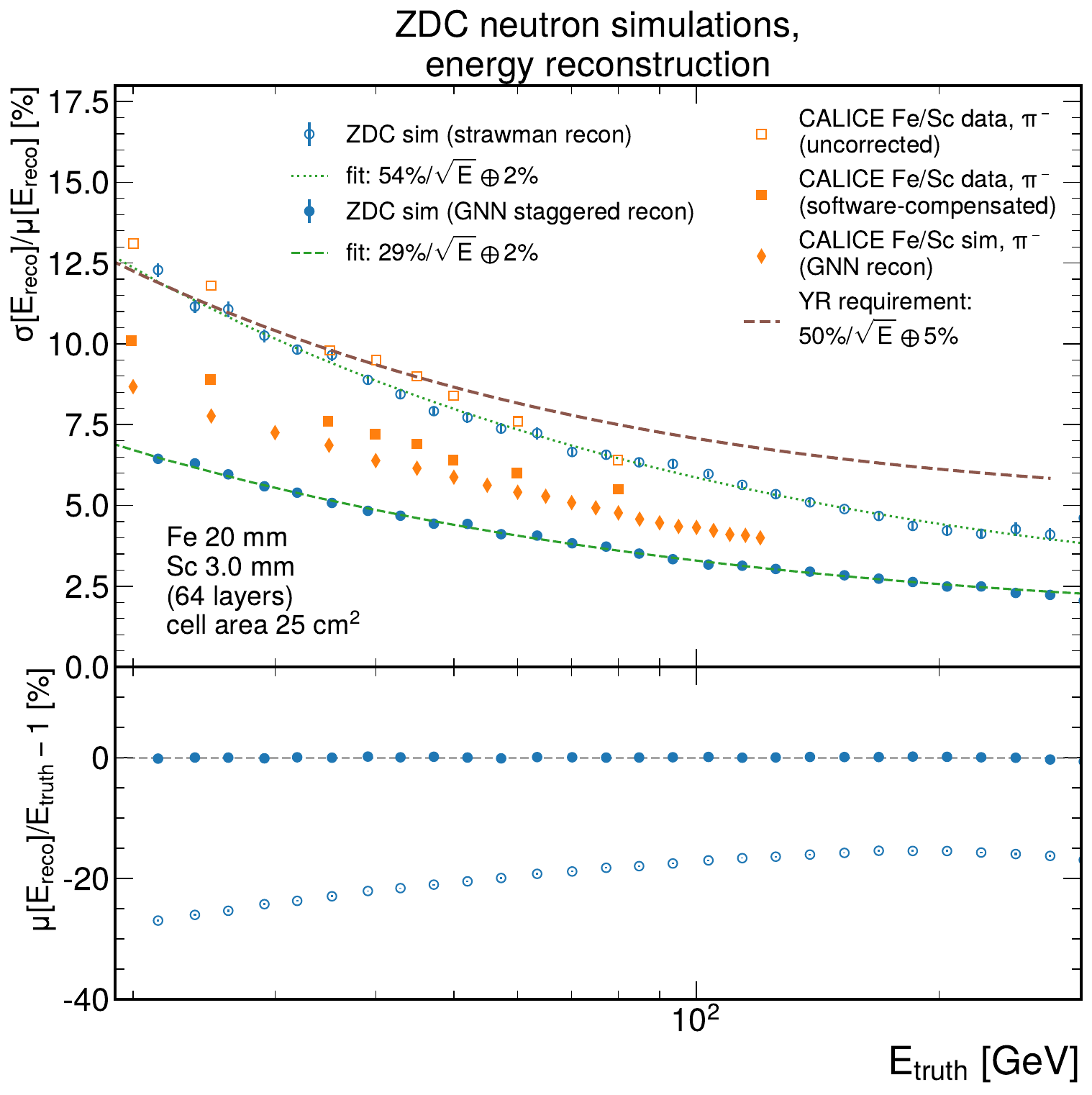}
    \caption{Energy resolution (top row) and scale (bottom row) obtained with the strawman (open symbols) and GNN (filled symbols) for simulated single neutrons. The resolutions are compared to those of the CALICE beamtest~\cite{CALICE:2012eac} (orange squares) with (filled) and without (open) software compensation, as well as the CALICE GNN reconstruction (orange diamond)~\cite{CALICE:2024imr}.}
    \label{fig:energy_resolution}
\end{figure}

We show the single-neutron energy resolution, $\sigma$, and scale for reconstructed neutrons from the strawman and GNN reconstruction algorithms in Fig.~\ref{fig:energy_resolution}.  In the scale the mean energy value, $\mu$, over the simulated events is compared to the true energy. We find that while the strawman reconstruction result has an offset in the scale up to $\approx-$30\%, the deviation from the perfect scale, 1, in the GNN reconstruction is nearly zero. The strawman reconstruction also exhibits non-linearities as seen by the non-flat behavior in the scale. When $E_{truth}$ vs. $\mu$ is fit with a linear function, we found deviations from linearity of 1-2\% for the strawman compared to nearly 0\% for the GNN, except at the lowest energies where it is between 0\% and 1\%. The resolutions for the strawman method are close to the $50\%/\sqrt{E}\oplus5\%$ requirement of the YR at low energy, and are below the YR requirement at higher energy.  The resolutions from the GNN method are considerably better, especially at lower energies.  

We compare the results of our simulation to those of a CALICE beamtest~\cite{CALICE:2012eac}, which included software compensation in the reconstruction.  We find that the resolutions from our strawman reconstruction are comparable to those of CALICE without software compensation. The resolution in our simulations with the GNN method outperforms that of CALICE with software compensation by about 30\%. We also compare to the CALICE GNN studies~\cite{CALICE:2024imr}, which our GNN method outperforms by around 30\%. Despite this, both GNN approaches perform better than the other reconstruction methods. Since the CALICE simulation has more realistic aspects such as a noise model than ours, the CALICE GNN results can be viewed as the lower limit of energy resolution for the ZDC.

We likewise measured the energy resolution for single-photon showers and found it to be $20\%/\sqrt{E}$ with an energy scale at unity for both the strawman reconstruction and GNN method, as expected.  

\begin{figure*}[h!]
    \centering
    \includegraphics[width=\textwidth]{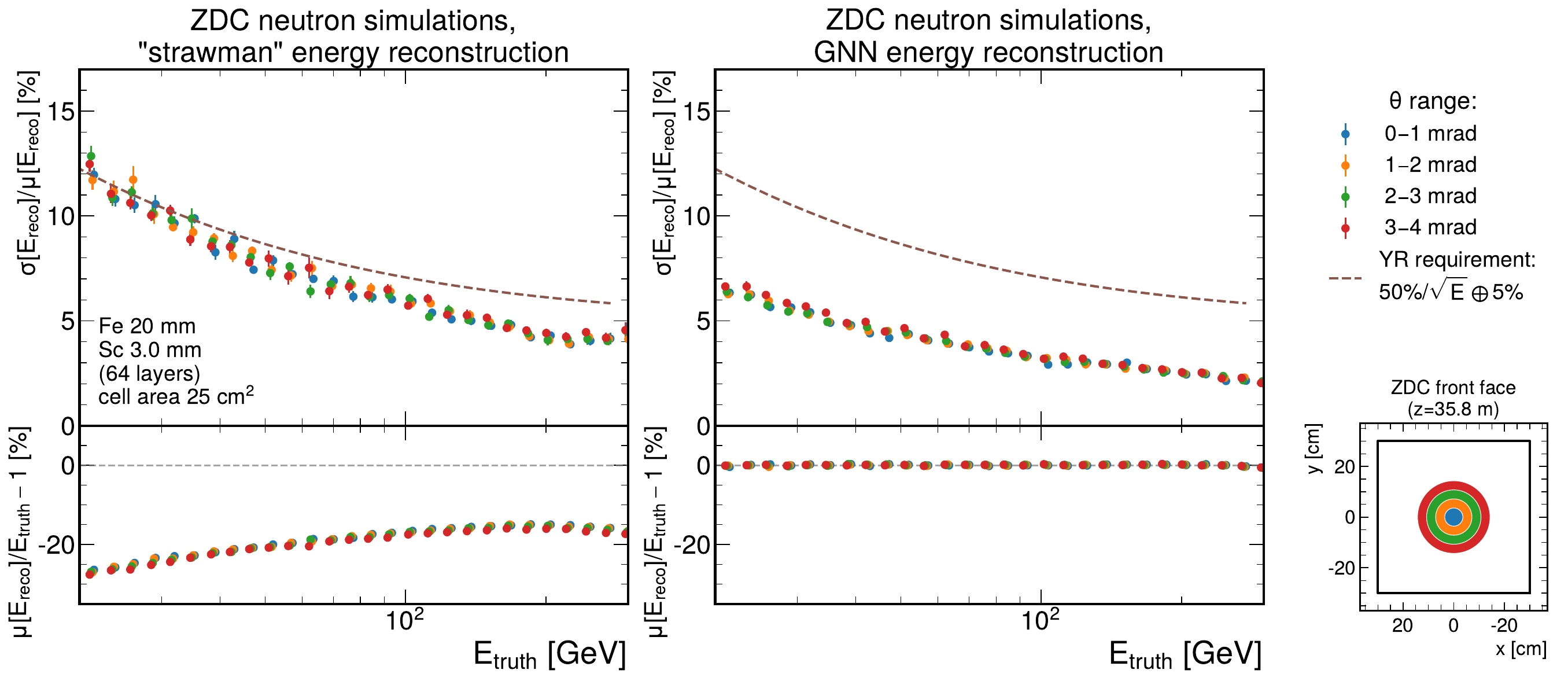}
    \caption{Energy resolution (top row) and scale (bottom row) obtained with the strawman (left) and GNN (right), at various ranges in $\theta$.  The positions on the face of the detector corresponding to each color are visualized in the inset.}
    \label{fig:energy_resolution_edge}
\end{figure*}

To determine if there were any edge effects within the fiducial region of the detector, we show the resolution and bias for the strawman and GNN reconstruction at various ranges in the polar angle $\theta$ in Fig.~\ref{fig:energy_resolution_edge}.  We find that the resolution does not have a strong dependence on the polar angle within this acceptance region.

We also compared the energy resolutions using the GNN method between the staggered and unstaggered layouts and found similar performance.

\subsection{Position resolution}
\begin{figure}[h!]
    \centering
    \includegraphics[width=0.495\textwidth]{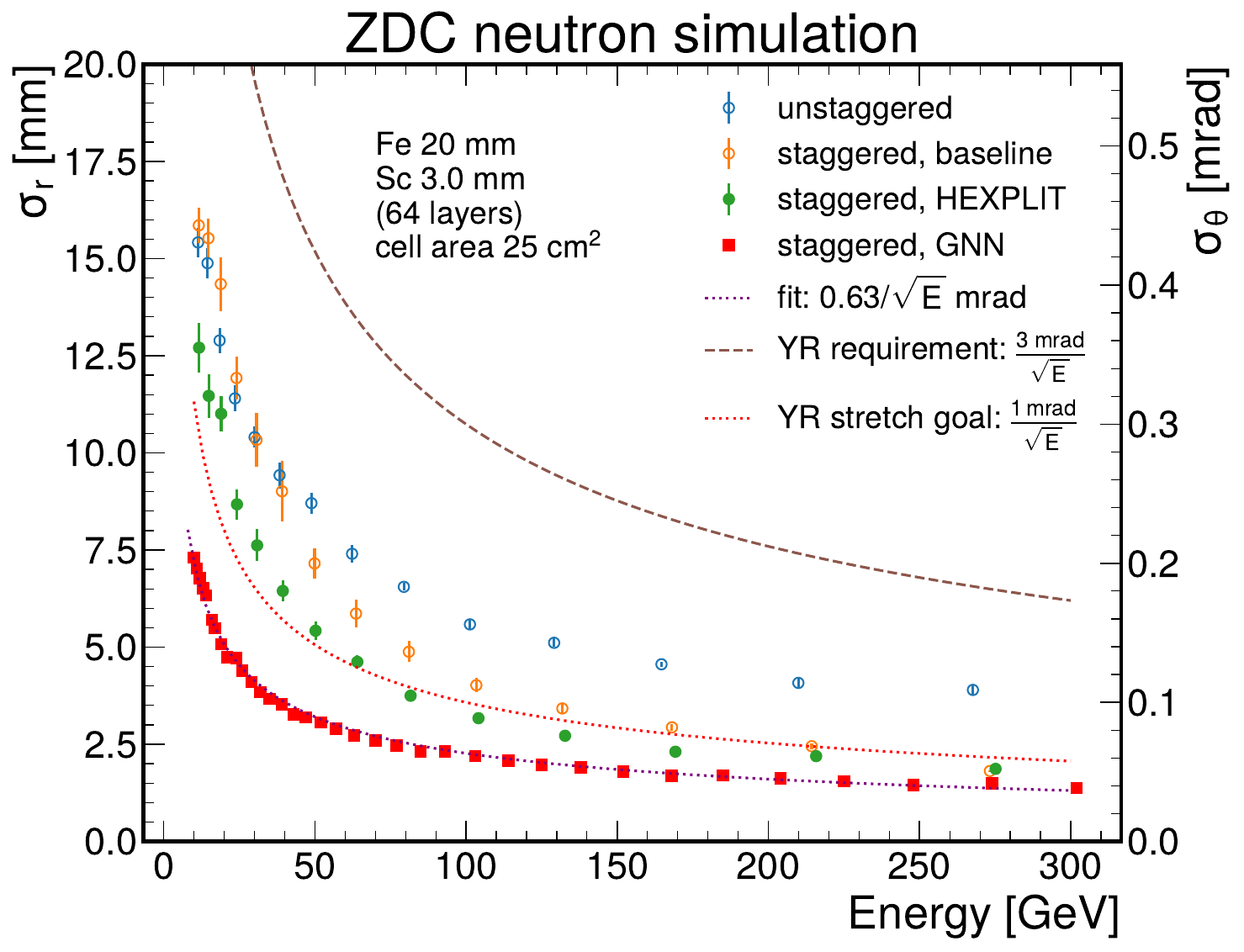}
    \caption{Position resolution for neutrons as a function of the generated energy.   Results are shown with an unstaggered layout (blue) and staggered layout with the positions reconstructed with the baseline (orange), HEXPLIT (green), and GNN (red) reconstruction algorithms.}
    \label{fig:position_resolution}
\end{figure}

\begin{figure}[h!]
    \centering
    \includegraphics[width=0.495\textwidth]{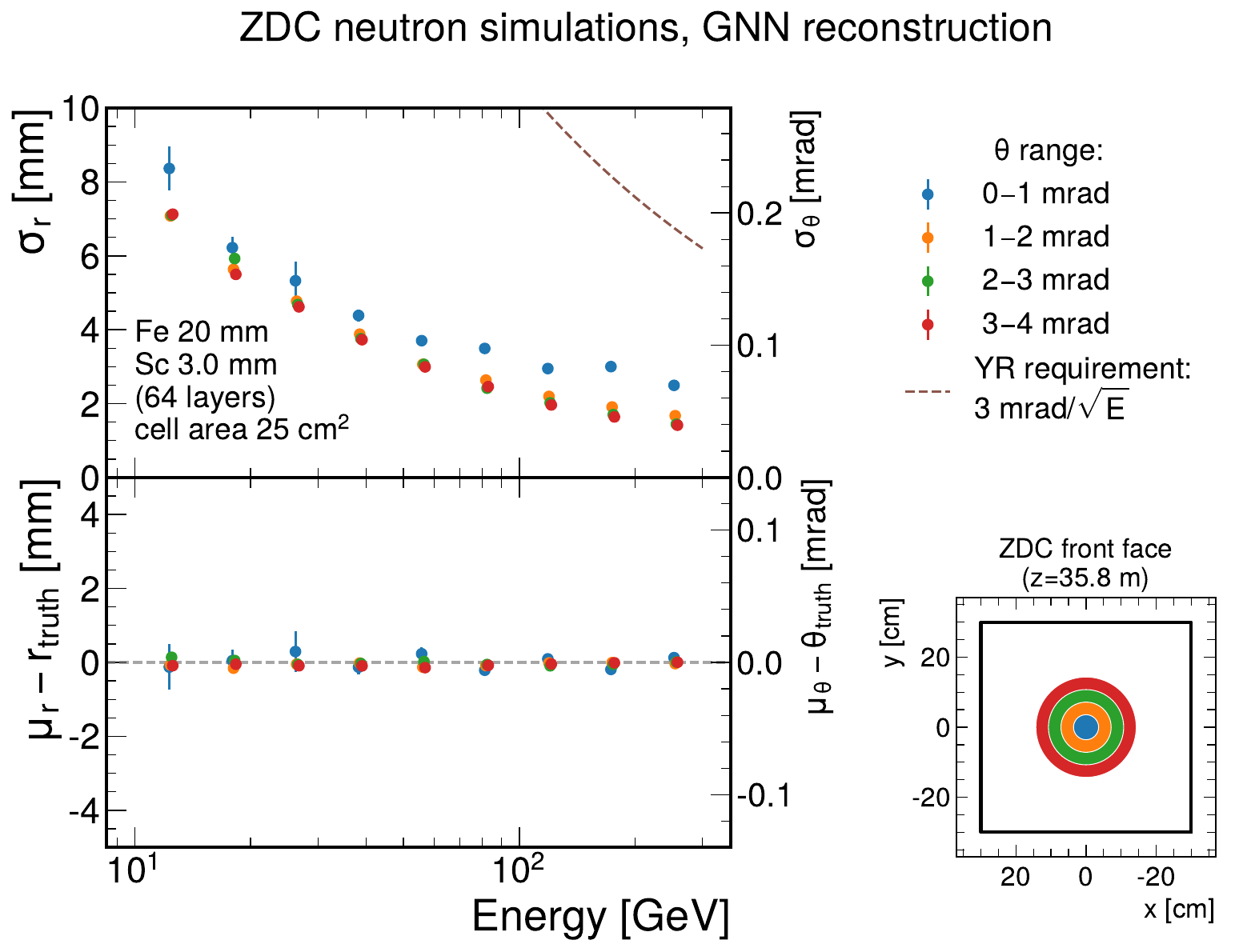}
    \caption{Top: same as Fig.~\ref{fig:position_resolution}, except in slices of $\theta$ and only for the GNN method.  Bottom: bias in position reconstruction for neutrons as a function of the generated energy.  The colored regions in the inset provide a visualization of where these slices of $\theta$ intersect the front face of the detector.}
    \label{fig:position_bias}
\end{figure}

Figure~\ref{fig:position_resolution} shows the position resolution as a function of the generated energy of simulated neutrons, while Figure~\ref{fig:position_bias} presents it in slices of polar angle.  This resolution is defined as the sigma of a Gaussian fit to the distributions of the radial position residuals,
defined by
\begin{equation}
    \Delta r= r_{\rm reco}-r_{\rm truth},
\end{equation}
where $r_{\rm reco}$ is the radial coordinate of the reconstructed position of where the particle struck the face of the detector, $r_{\rm truth}$ is the radial coordinate of the position of the truth particle track at the detector face. The angular resolution is then the radial position resolution divided by the distance from the nominal interaction point to the front face of the detector.  

We compared the results obtained with the staggered layout proposed in this work to those obtained with an unstaggered layout. Further, we compare the resolutions obtained with the ``baseline'' and HEXPLIT reconstruction algorithms, as described in Ref.~\cite{hexplit}, and the GNN reconstruction.  We find that the best resolution is obtained using the GNN. 
 At high energies, the HEXPLIT and GNN algorithms produce nearly the same resolutions as one another, while at lower energies, the HEXPLIT performs worse. The staggered-layer design and GNN reconstruction easily meet the requirements outlined in the EIC Yellow Report~\cite{EICYR} and are close to the more stringent requirement set forth in Ref.~\cite{Bylinkin:2022rxd}.  

At 100 GeV, the angular resolution with the GNN reconstruction is 63 $\mu$rad, which, when added in quadrature with the beam divergence of 56 $\mu$rad in the high-acceptance configuration~\cite{Bylinkin:2022rxd}, corresponds to a $p_T$ resolution of 8.4~MeV.

We repeated this exercise for single photons and obtained a resolution of $\sigma_\theta=0.19/\sqrt{E}\oplus 0.014$~mrad for the baseline algorithm. 
 For reference, the smallest separation between photons produced in $\pi^0$ decay allowed by kinematics is 1.0 mrad, about an order of magnitude larger than the single-photon position resolution. We discuss $\pi^0$ and $\gamma$ classification in Section~\ref{sec:pi0}. 

\subsection{Multiple-neutron events}
\begin{figure}[h!]
    \centering
        \includegraphics[width=.42\textwidth]{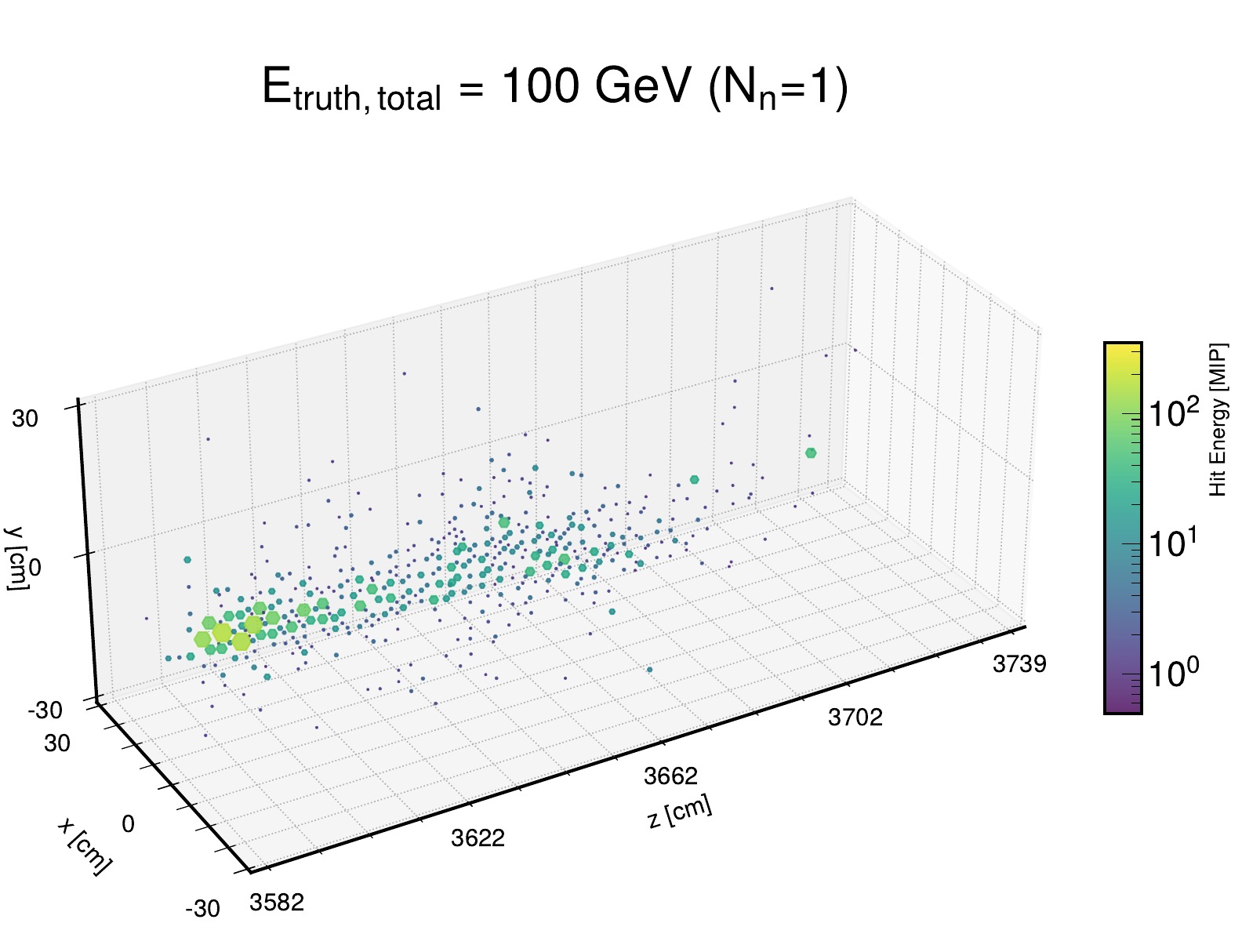}
        \includegraphics[width=.42\textwidth]{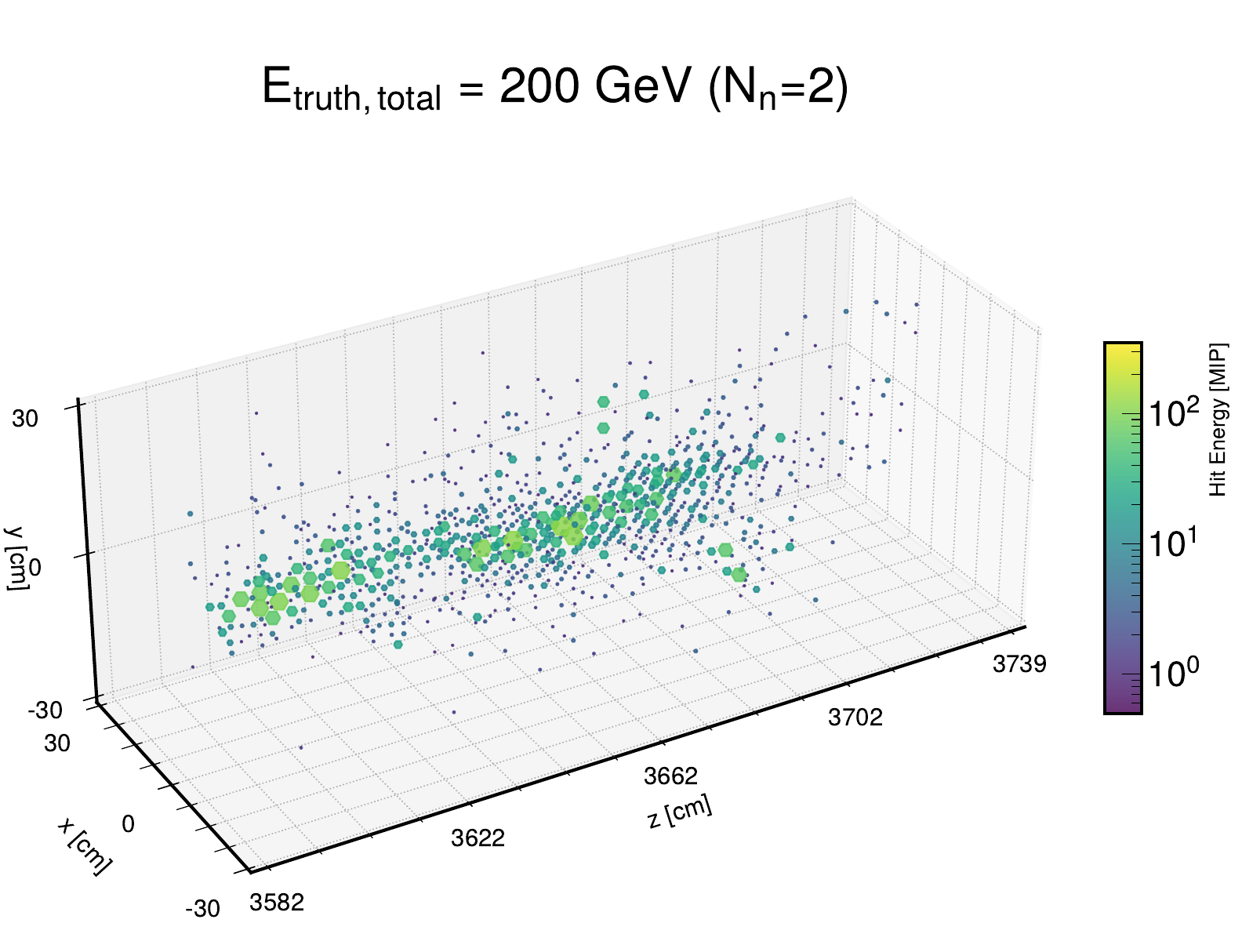}
        \includegraphics[width=.42\textwidth]{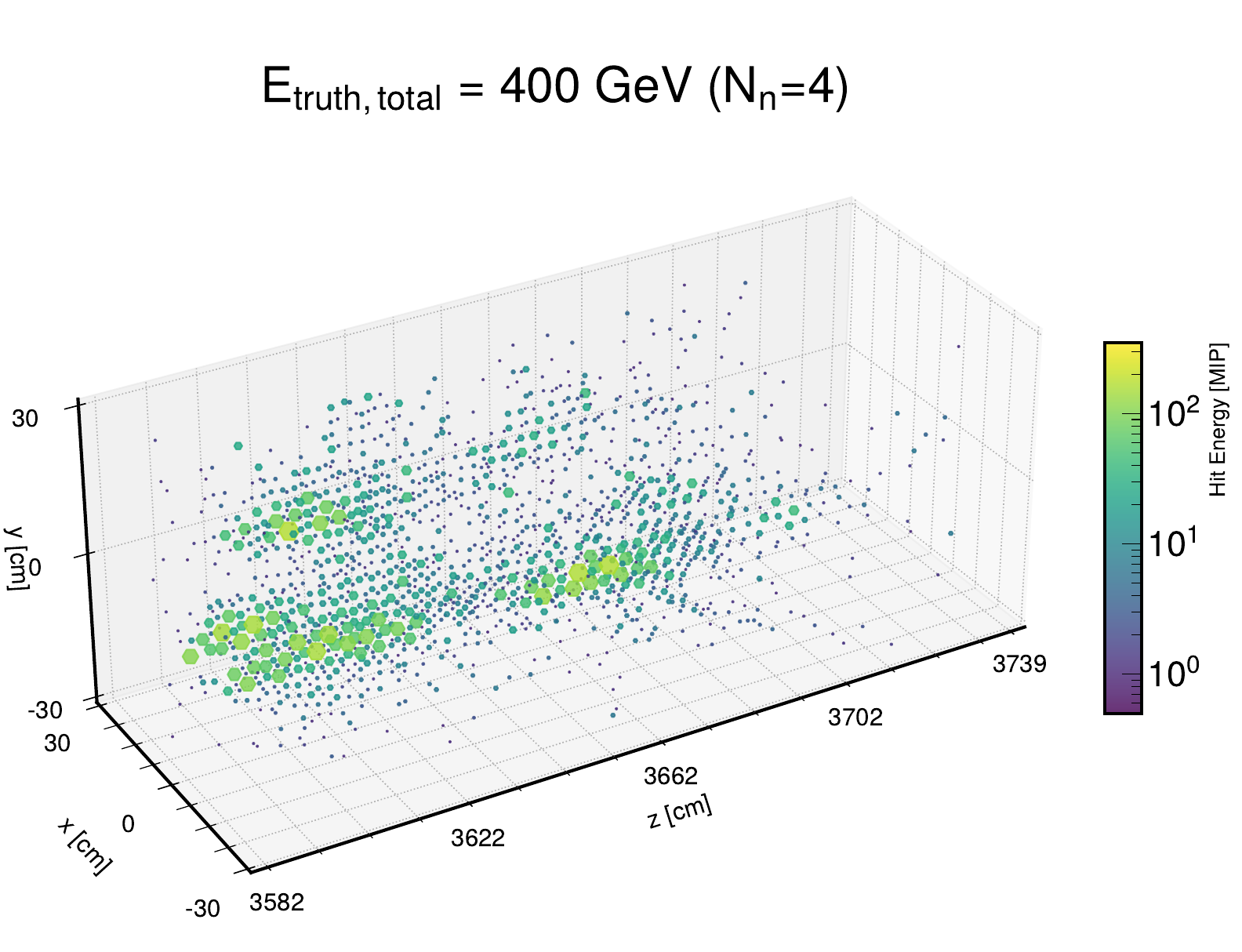}
        \includegraphics[width=.42\textwidth]{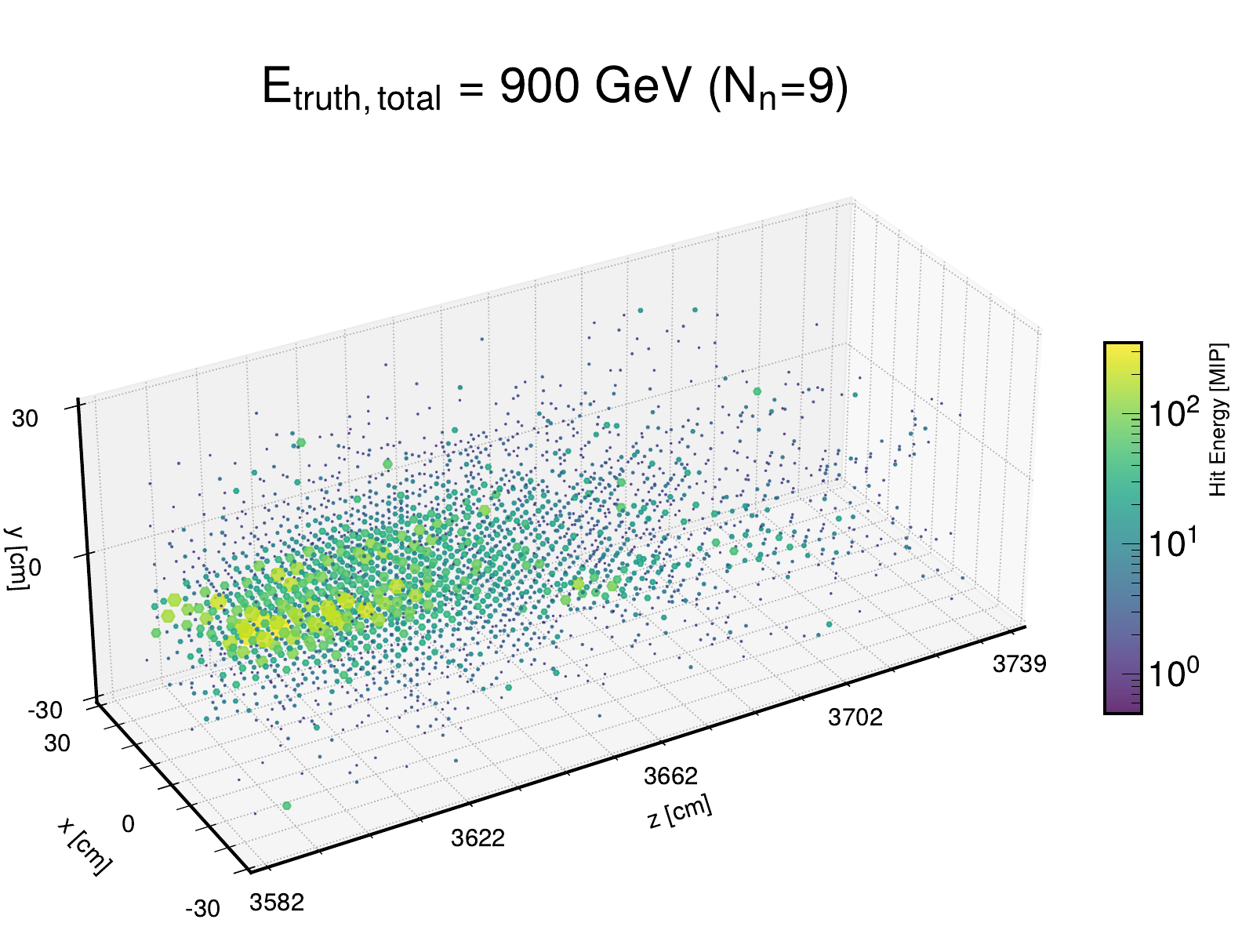}
        \caption{Examples of 4 reconstructed 3D shower shapes in the ZDC for events with 1 neutron ($N_{n}=1$), 2 neutrons ($N_{n}=2$), 4 neutrons ($N_{n}=4$), and 9 neutrons ($N_{n}=9$). The color code represents hit energy in terms of $E_\mathrm{MIP}$. The marker size is displayed proportionally to hit energy for display purposes. }
    \label{fig:multineutron_showers}
\end{figure}

We evaluate the ZDC performance for multiple-neutron events (illustrated in Fig.~\ref{fig:multineutron_showers}) that are expected in $eA$ collisions, where the spectator neutrons mostly have the same energy modulo ``nuclear effects'' such as Fermi motion, short-range correlations or fission\footnote{
Upon boosting to the laboratory frame, these nuclear effects led to a Gaussian smearing with a width of about 5 GeV around the beam momentum of 100 GeV per nucleon, as per the \textsc{BeAgle} event generator~\cite{Chang:2022hkt}.}
.

\begin{figure}[h!]
    \centering
      \includegraphics[width=\linewidth]{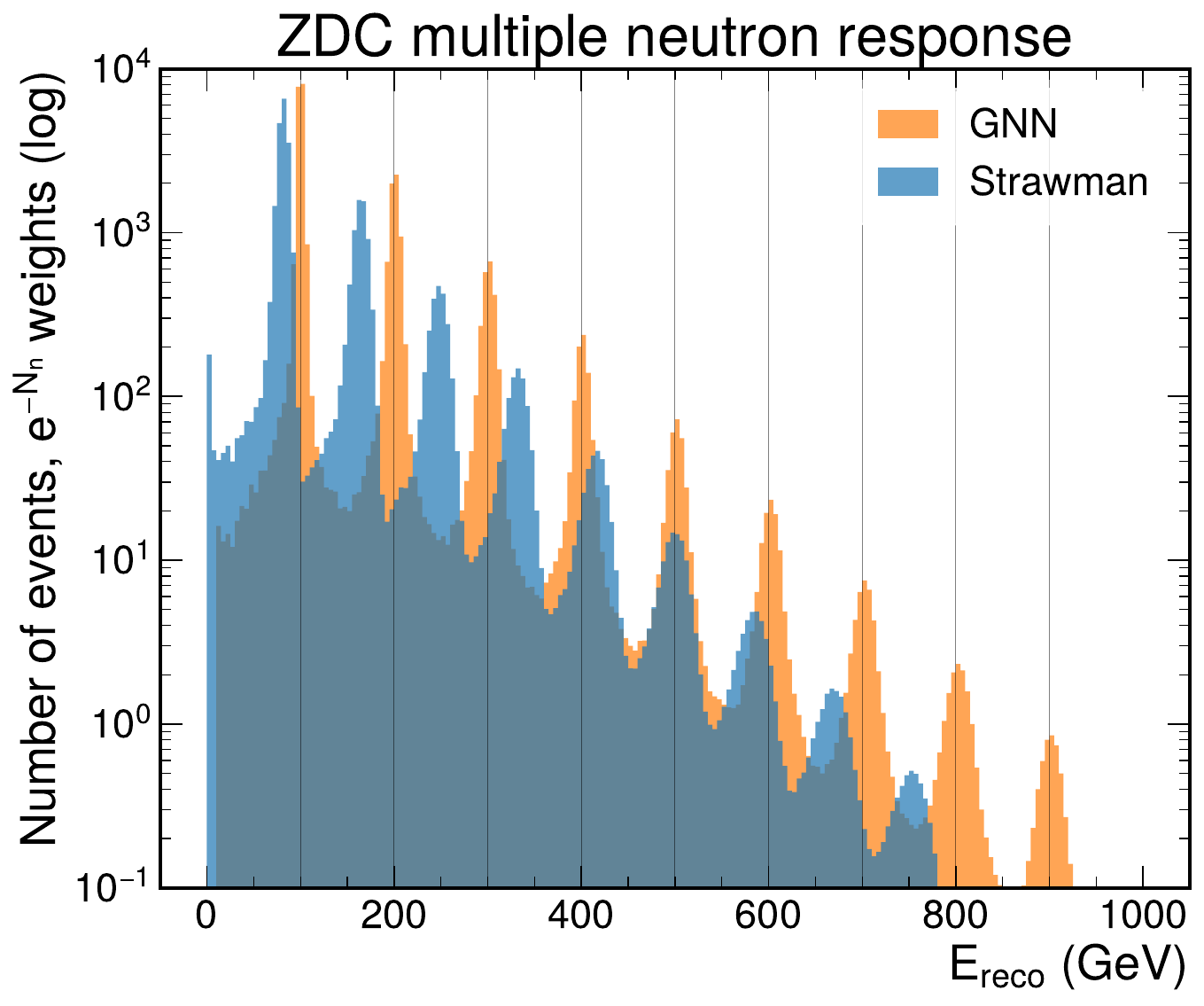}
      \captionof{figure}{Energy measured for each integer number of neutrons with energy equal to 100 GeV, reconstructed with GNN (orange) and strawman (blue).  The events are weighted by a factor of $e^{-N_{n}}$. The black, vertical lines are drawn at the total true energies of events with different numbers of neutrons. Here we show events containing up to nine neutrons. This performance illustrates just instrumental effects, excluding nuclear effects like smearing caused by Fermi motion and fission. In the laboratory frame, these nuclear effects would contribute in quadrature about 5 GeV per 100 GeV neutron, according to \textsc{BeAgle}~\cite{Chang:2022hkt}, which happens to be similar to the widths of the strawman reconstruction shown here.
     }
      \label{fig:multipleneutron_spectrum}
\end{figure}

Figure~\ref{fig:multipleneutron_spectrum} 
compares the multi-neutron performance of the strawman approach and the GNN at a neutron energy of an integer multiple of exactly 100 GeV and a random angle within 4 mrad; we excluded the anticipated smearing from nuclear effects in assessing this performance to isolate purely instrumental effects. 

The reconstructed energy peaks for each number of neutron ($N_{n}$) are clearly separated in both approaches. Similar to the single-neutron case, the GNN adjusts the energy scale and restores the linearity and reduces the width of the peaks compared to the strawman reconstruction. This improvement occurs despite the overlapping showers and event complexity illustrated in Fig.~\ref{fig:multineutron_showers}. 

The energy resolutions for these multiple-neutron events are worse than the performance extrapolated from single-neutron events (shown in Fig.~\ref{fig:energy_resolution}), by about 10--30\% in the range studied; for instance, the width for five 100 GeV neutrons is about 8 GeV, whereas extrapolating from single-neutron performance one would estimate it to be about 7 GeV. This degradation might originate from the overlap of showers, potentially resulting in a loss of information about the local densities of individual showers.

The reconstructed peak widths in neutron measurements in $eA$ collisions, similar to Ref.~\cite{ALICE:2022iqi}, are expected to receive comparable contributions from instrumental effects and nuclear effects when using the strawman reconstruction, and to be dominated by nuclear effects when using the GNN method.

\FloatBarrier

\subsection{$\pi^0/\gamma$ separation}
\label{sec:pi0}
\begin{figure}[h!]
    \centering
      \includegraphics[width=\linewidth]{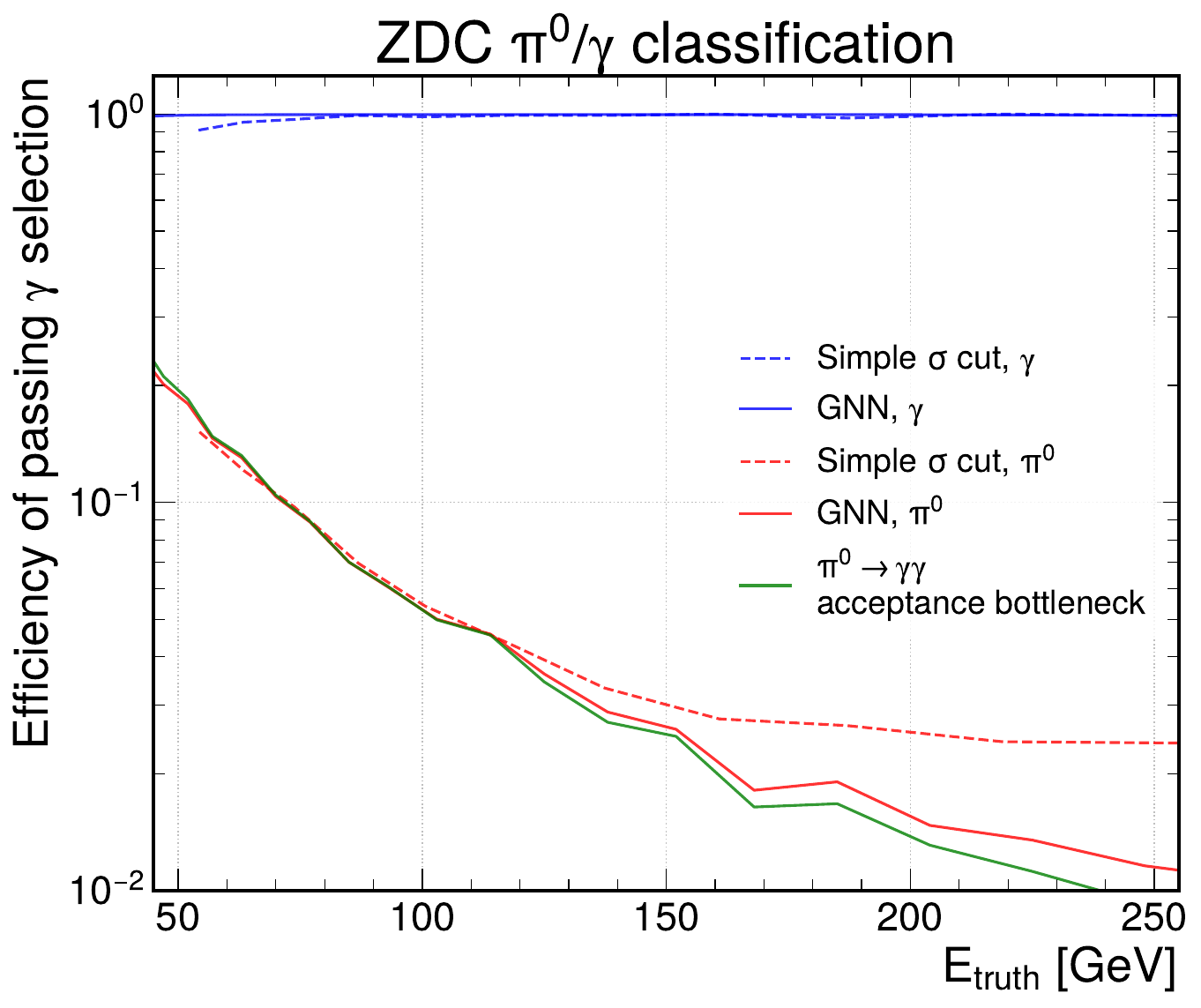}
      \captionof{figure}{Efficiency of classifying $\gamma$ (blue) and $\pi^0$ (red) as $\gamma$. The $\pi^0$ rejection efficiency has a bottleneck (green) at lower energies due to events with only one photon hitting the ZDC. The green curve is  the fraction of events where only one decay photon hits the ZDC, and is therefore a bottleneck on separating a $\gamma$ from a $\pi^0$.}
      \label{fig:efficiency}
\end{figure}

We quantify the model's ability to separate $\pi^0$ and $\gamma$ using the efficiency, as shown in Fig.~\ref{fig:efficiency}. This shows the fraction of events that the model classified as photons, \textit{i.e.}, the fraction of events with outputs below the classification cut of 0.3. The GNN consistently classifies photon events with $99\%$ efficiency between 50 and 250 GeV, and correctly classifies $98\%$ of $\pi^0$ events above 150 GeV. 

At lower energies, the GNN misclassifies up to $20\%$ of $\pi^0$ events. At these energies, the separation between photons is large enough such that often only one photon deposits energy in the ZDC in some events, making it impossible for the GNN to differentiate these events from single-photon events. As such, the GNN offers similar separation to that from a simple $\sigma$ cut on the shower width, indicating no substantial improvement can be achieved in $\pi^0/\gamma$ separation at low energies. When both photons consistently reach the ZDC at higher energies, the $\sigma$ cut and GNN achieve a $\pi^0$ rejection efficiency greater than $97\%$ and $98\%$, respectively.

\FloatBarrier
\section{Summary}
\label{sec:summary}

We have presented a design and simulated the performance of a high-granularity Zero-Degree Calorimeter for the future Electron-Ion Collider. This design is similar to the CALICE AHCAL prototypes and  uses iron blocks as absorbers, with scintillator readout using SiPMs. Unlike CALICE designs, it includes a novel staggered layer design based on hexagonal tessellation patterns, aimed at enhancing its position resolution.

We demonstrate the efficacy of this design using a machine-learning regression based on a graph representation, which provides motivation for choosing this design over traditional low-granularity ones. We show that the GNN can effectively handle and exploit the complex pattern of staggered hexagonal cells for software compensation and particle identification. The GNN improves energy reconstruction, enhancing resolution and scalability compared to other approaches for both single-neutron events and multiple-neutron events. It also provides performance for single photon vs. neutral pion classification that is close to optimal given its acceptance.

The current design meets the expectations outlined in the EIC Yellow Report, especially in terms of improving angular resolution. Additionally, its high granularity enhances background-rejection capabilities, such as beam-gas interactions and SiPM noise. This design could also offer fine time resolution, which could be used to improve energy regression or background rejection. We will explore this in future work after prototype beam testing reveals realistic time performance.

This ZDC design will be capable of delivering performance for the majority of the physics program at EIC, which necessitates measurements of high-energy neutrons, photons, and neutral pions. Additionally, this design could be complemented with a homogeneous crystal calorimeter designed for measuring low-energy $O(10-100)$ MeV photons, which would serve the purpose of tagging backgrounds for coherent $eA$ scattering~\cite{Chang:2021jnu}, and other applications. 

The design and GNN-based reconstruction presented in this work can serve as a blueprint for guiding the design and optimizing performance for other high-granularity calorimeter systems at the EIC. These include the forward hadronic calorimeter~\cite{ECCE_CALO}, and especially its high-granularity insert~\cite{Insert}, the barrel electromagnetic and hadronic calorimeters, the few-degree calorimeter~\cite{Arratia:2023gyx}, among others. 

\section*{Code Availability}
The code for the data processing, training models, and plotting results can be found here: \url{https://github.com/eiccodesign/regressiononly/tree/zdc_classification}. 
The data used in these studies are found in Refs.~\cite{milton_2024_11177058,milton_2024_11177121,milton_2024_11176795,milton_2024_11177528,milton_2024_11177711,milton_2024_11179550,milton_2024_11177424}. The GNN models stored in the ONNX format are found in Ref~\cite{milton_2024_11187659}.
 \section*{Addendum}
While this manuscript was in preparation, this ZDC design was incorporated into the baseline design of ePIC, which is the EIC project detector.
 \section*{Acknowledgments}
 We thank members of the California EIC consortium for their feedback on our design, especially Oleg Tsai. Additionally, we extend our gratitude to the ePIC collaboration, specifically Alexander Jentsch and Elke-Caroline Aschenauer, for their numerous discussions about ZDC physics and detector design.
 
 We acknowledge support from DOE grant award number DE-SC0022355.  We also acknowledge support by the MRPI program of the University of California Office of the President, award number 00010100. S.P also acknowledges support from the Jefferson Lab EIC Center Fellowship. This research used resources from the LLNL institutional Computing Grand Challenge program and the National Energy Research Scientific Computing Center, a DOE Office of Science User Facility supported by the Office of Science of the U.S. Department of Energy under Contract No. DE-AC02-05CH11231 using NERSC award HEP- ERCAP0021099. M.A acknowledges support through DOE Contract No. DE-AC05-06OR23177 under which Jefferson Science Associates, LLC operates the Thomas Jefferson National Accelerator Facility. This work was performed under the auspices of the U.S. Department of Energy by Lawrence Livermore National Laboratory under Contract No. DE-AC52-07NA27344.

\FloatBarrier
\renewcommand\refname{Bibliography}
\bibliographystyle{utphys} 
\bibliography{bibio.bib} 

\appendix
\end{document}